\documentstyle[12pt,epsfig,docbib]{article}
\renewcommand{\vec}[1]{\relax\ifmmode\mathchoice
{\mbox{\boldmath$\relax\displaystyle#1$}}
{\mbox{\boldmath$\relax\textstyle#1$}}
{\mbox{\boldmath$\relax\scriptstyle#1$}}
{\mbox{\boldmath$\relax\scriptscriptstyle#1$}}\else
\hbox{\boldmath$\relax\textstyle#1$}\fi}
\newcommand{\abbl}[4]{%
\begin{figure}[htbp]\vspace*{0.5cm}%
\parbox[b]{6cm}{\begin{center} #1 \end{center}}#4\hfill%
\parbox[b]{9cm}{%
\caption[#2]{#3\\ \mbox{ }}
}%
\vspace*{0.5cm}\end{figure}}        
\begin{document}
\pagestyle{myheadings}
\markboth{Helbing/Vicsek: Self-Organized Optimality}
{Helbing/Vicsek: Self-Organized Optimality}
\title{{\bf Optimal Self-Organization}}
\author{\normalsize
Dirk Helbing$^{*,+}$ and Tam\'{a}s Vicsek$^+$\\[4mm] 
\normalsize {\tt http://www.theo2.physik.uni-stuttgart.de/helbing.html}\\[-1mm]
\normalsize and {\tt http://angel.elte.hu/$\widetilde{\hphantom{I}}$vicsek/ }}
\renewcommand{\today}
{\normalsize $^*$ II. Institute of Theoretical Physics, University of Stuttgart,\\
\normalsize Pfaffenwaldring 57/III, 70550 Stuttgart, Germany\\[4mm]
\normalsize $^+$ Department of Biological Physics, E\"otv\"os University, Budapest,\\
\normalsize P\'azm\'any P\'eter S\'et\'any 1A, H-1117 Hungary}
\maketitle
\hfill
\begin{abstract}
{\normalsize We present computational and analytical results indicating
that systems of driven entities with repulsive interactions
tend to reach an optimal state 
associated with minimal interaction and minimal dissipation.
Using concepts related to those from non-equilibrium 
thermodynamics as well as game theoretical ideas, 
we generalize this finding 
to an even wider class of self-organizing systems  
which have the ability to reach a state of maximal overall ``success''. 
This principle is expected to be relevant
for driven systems in physics like sheared granular media, 
but it is also applicable to 
biological, social, and economic systems, for which only a limited number
of quantitative principles are available yet.}
\end{abstract}
\hfill
\pagebreak

\section{Introduction}\label{intro}

Extremal principles
are fundamental in our interpretation of phenomena in 
nature. One of the best known examples is the second law of
thermodynamics \cite{Keizer,Info,Pri}, 
governing most physical and chemical systems
and stating the continuous increase of entropy (``disorder'')
in closed systems. Most systems in our natural environment, however, are
open, which is true for driven physical systems, but even more
for biological, economic, and social systems. As a consequence,
these systems are usually characterized by self-organized structures 
\cite{Info,Pri,Haken,evol,Ebeling,gamedyn,quantsoc,frank,eshel,biol1a,biol1b,%
biol1c,biol2,biol3,biol4,biol5,biol6,Drasdo,ants1,ants2,mandel,Bouchaud,solomon,stanley,peinke,%
zhang,Stauf,We91,Hub,Hel93,Galam,pre,granular,solid,trail},
which calls for principles that apply
on time scales shorter or comparable to the life spans of these systems.
For example, it is known that in growth and aggregation processes
it is usually the most unstable mode that determines the finally evolving
structure. Hence, there exists an extremum principle of fastest propagation,
which is applicable to so apparently different phenomena like
crystall growth on the one hand \cite{jacob1} and pattern formation in 
bacterial colonies on the other hand \cite{jacob2}. 
\par
\begin{figure}[htbp]
\begin{center}
\epsfig{width=12.0\unitlength, angle=-90,
      file=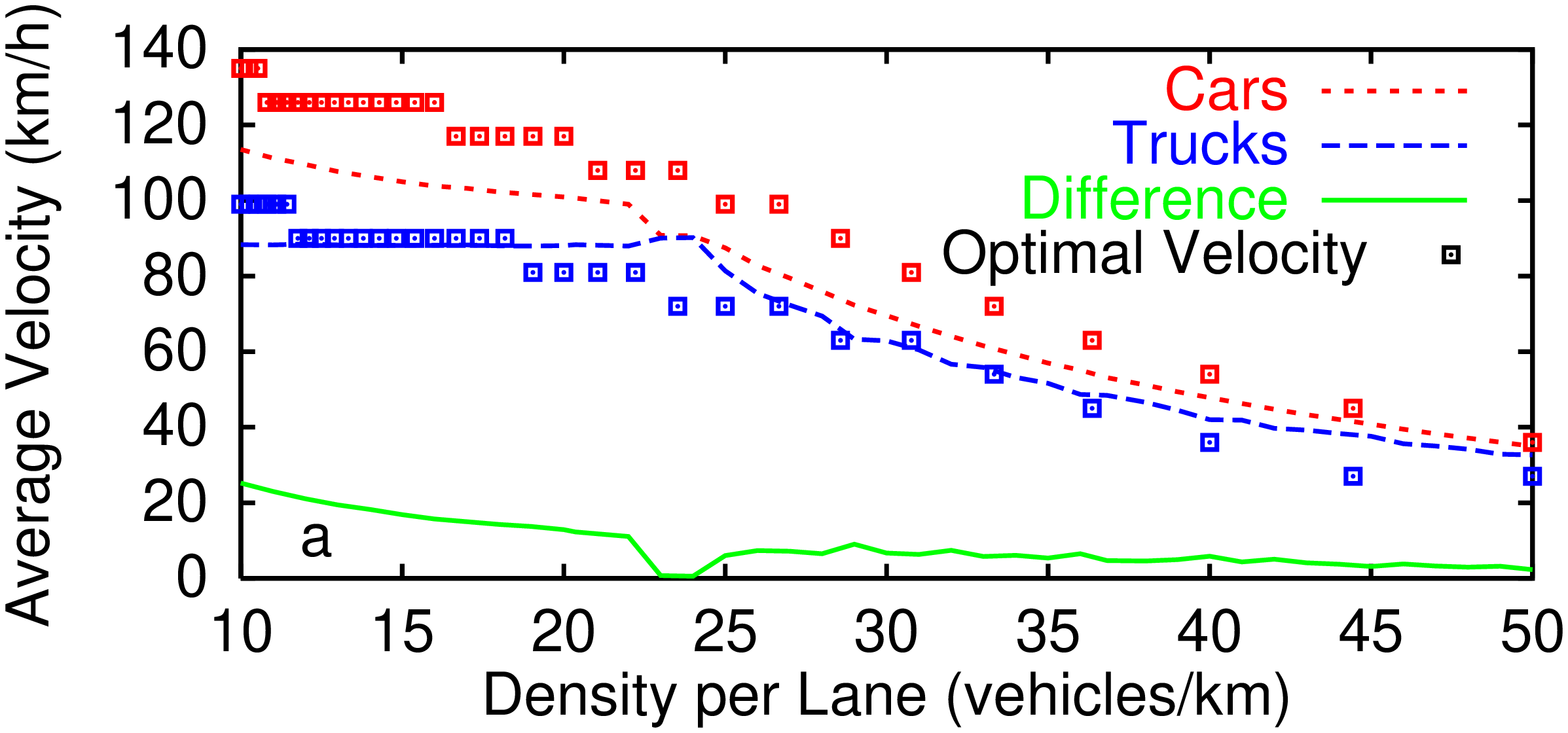} 
\vspace*{-2\unitlength}
\epsfig{width=12.0\unitlength, angle=-90,
      file=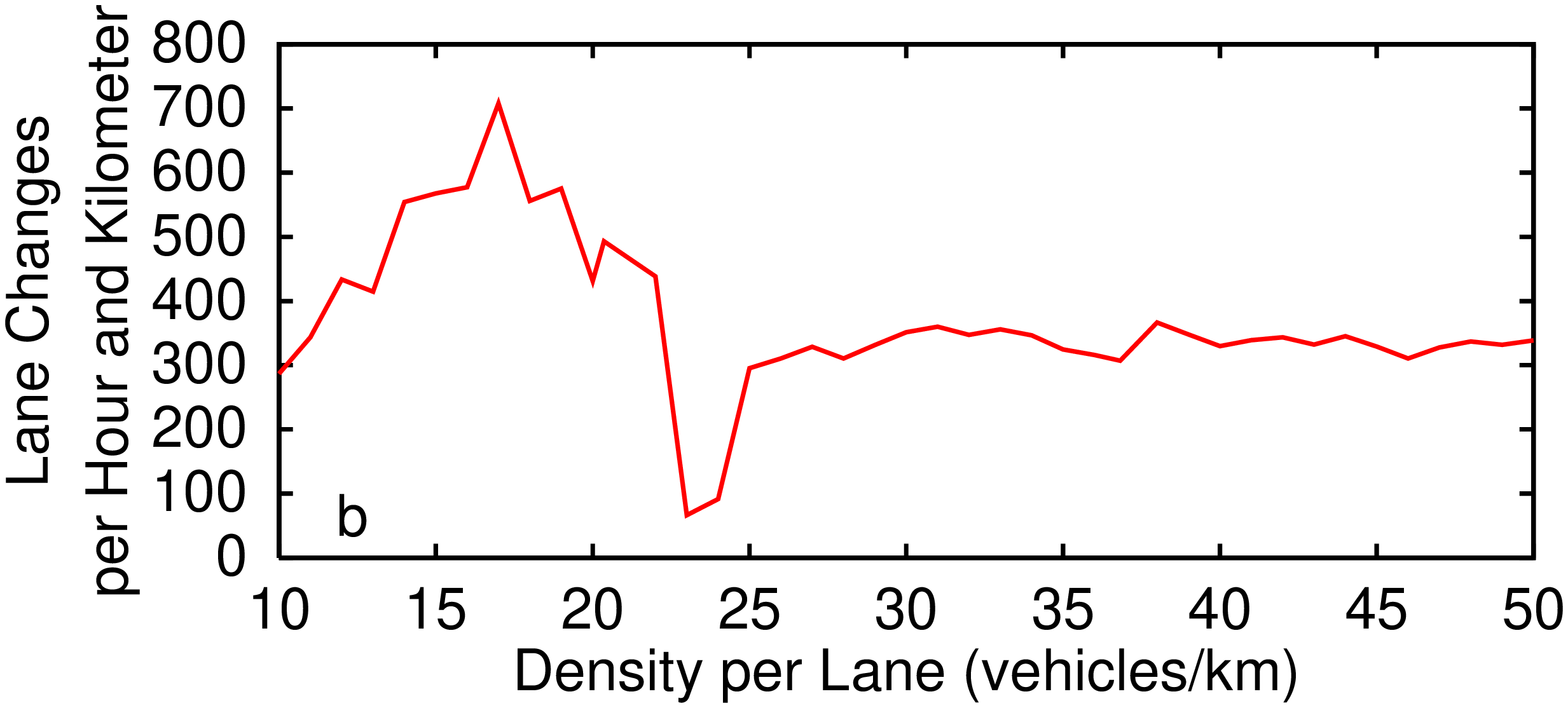} 
\end{center}
\vspace*{6mm}
\caption[]{Simulation results for a uni-directional two-lane freeway
used by two different vehicle types, cars and trucks.
We assume that fast vehicles overtake slow ones,
if possible and safe. 
In addition, cars tend to drive with their respective
``optimal (safe) velocity'' (red squares),
which depends on the local vehicle density (the
inverse distance to the next vehicle ahead). The same applies to
trucks, but with a considerably smaller optimal velocity (blue squares).
(For details see Ref.~\cite{solid}.)
According to our
simulations, traffic flow is stable up to densities of about 24 vehicles per
kilometer and lane, while stop-and-go traffic develops at higher densities.
{\bf a} Between about
22.5 and 24 vehicles per kilometer and lane, the difference (green line)
between the average velocities of cars (red line) 
and trucks (blue line) becomes almost zero, which is a consequence of the 
breakdown of the lane changing rate (see {\bf b}). Hence, a 
coherent state of motion appears only in a small density range
which, at the same time, is characterized by a minimal interaction rate and
a minimal overtaking rate. Hence, self-organization 
(coherent motion) and optimality (minimal interaction) are directly 
related in this example.\label{sol}}
\end{figure}

Recent simulations point to the possible existence of
additional optimality principles in certain kinds of
driven multi-particle or multi-agent systems. As examples we mention
\begin{enumerate}
\item
lane formation in pedestrian crowds \cite{pre} (see Fig.~\ref{fig2}),
which appears to be similar to the size segregation in 
sheared granular media \cite{granular}, 
\item
the self-organization of coherent motion in a mixture of cars and trucks
\cite{solid} (see Fig.~\ref{sol}), and
\item the evolution of trail systems \cite{trail} (see Fig.~\ref{tr}).
\end{enumerate}
In these systems, the respective interacting entities (pedestrians,
driver-vehicle units, or
particles) have to coordinate each other in order to reach a 
system state which is ``favourable'' to them.
It was conjectured \cite{trail,solid} that the resulting system states are 
optimal in some sense, but
there are many open questions to be addressed:
\begin{enumerate}
\item Is there really a quantity which is optimized by the
self-organizing system in the course of time?
\item If yes, which quantity is it? Is there a systematic way to 
derive it?
\item What are the conditions for the existence of such a quantity?
\item Is there any systematic connection between self-organization 
and optimization?
\item If optimal self-organized systems exist at all, are they
exceptional or quite common?
\end{enumerate}
We will address these questions in the following sections. A particular
problem will be, how to formulate the theoretical approach to the problem
general enough to grasp the variety of different systems mentioned
above. We will manage this by applying concepts related to those from
thermodynamics 
\cite{Keizer,Info,Pri,Ebeling} as well as game theoretical ideas
\cite{Ebeling,gamedyn,quantsoc,frank,game1,game2,game3,gd,Nowak,%
life,may,Bernardo}.

\section{An Example: Lane Formation}\label{laneform}

To illustrate the non-trivial aspects of optimal self-organization,
let us consider the dynamics of pedestrian crowds. We begin with a system of
oppositely moving pedestrians in a corridor, for which lane formation has been
observed in empirical studies \cite{book}. Readers who prefer
an example from physics may instead imagine a long vertical
column of a viscous fluid with
light (rising) and heavy (sinking) particles of equal size
(where we will assume that the absolute density difference of the 
particles with regard to the fluid is the same). In fact, we encourage
the reader to do this new experiment.
\par
By $\vec{x}_\alpha(t)$ we denote the position of pedestrian or 
particle $\alpha$
at time $t$, by $\vec{v}_\alpha(t) = d\vec{x}_\alpha(t)/dt$
its velocity, by $v_0$ its equilibrium speed in the absence 
of (interparticle) interactions, 
and by $\vec{e}_\alpha$ its ``desired'' or ``preferred'' direction of motion. 
Then, the equation mimicking pedestrian or describing particle motion
reads
\begin{equation}
 \vec{v}_\alpha(t) = v_0 \vec{e}_\alpha + \sum_{\beta (\ne \alpha)}
 \vec{f}_{\alpha\beta}\big(d_{\alpha\beta}(t)\big) 
\end{equation}
(in the overdamped limit). 
$\vec{f}_{\alpha\beta}$ represents repulsive interactions between
pedestrians or particles $\alpha$ and $\beta$, which were assumed to decrease
monotonically with their distance $d_{\alpha\beta}(t)=
\|\vec{x}_\alpha(t) - \vec{x}_\beta(t)\|$. We will not specify the exact form
of $\vec{f}_{\alpha\beta}$, since it turns out to be
quite irrelevant for the kind of phenomena 
we want to describe, here. The forces can be even chosen in a 
velocity-dependent way \cite{book,freezing}.

In cases of two opposite desired directions of motion, simulations of the 
above model reproduce the 
formation of lanes of uniform walking directions observed for pedestrians
(see Fig.~\ref{fig2}),
while there are no stable self-organized states in cases of four different
desired directions of motion (e.g., at intersections).\footnote{The reader
is invited to do his own simulation experiments with our Java Applets for 
lane formation and intersections available on this www-page:\\
{\tt http://www.theo2.physik.uni-stuttgart.de/helbing.html}}
It is clear that lane formation will maximize
the average velocity in the respective desired walking direction 
and, therefore, the quantity
\begin{equation}
 E(t) = \frac{\langle\langle \vec{v}_\alpha \cdot \vec{e}_\alpha
 \rangle_\alpha \rangle_t}{v_0} \le 1 \, , 
\end{equation}
which is a measure of the ``efficiency'' or ``success'' of motion.
(Here, $\langle\langle . \rangle_\alpha \rangle_t$ denotes the
average over the pedestrians and over time.) Moreover, optimization of
efficiency immediately implies that the system minimizes the quantity
\begin{equation}
\bigg\langle\!\bigg\langle - \sum_{\beta (\ne \alpha)}
 \vec{f}_{\alpha\beta} \cdot \vec{e}_\alpha 
 \bigg\rangle_{\!\!\alpha}\bigg\rangle_{\!\!t} 
 = v_0 - \langle\langle  \vec{v}_\alpha \cdot \vec{e}_\alpha
 \rangle_{\alpha} \rangle_{t} = v_0 (1 - E) \, ,
\end{equation}
i.e., the average interaction intensity 
opposite to the respective desired direction of motion.

Hence, even without the use of difficult mathematics we could show that
{\em the system minimizes the interaction intensity}
of the pedestrians (or particles), if it shows segregation into lanes
of uniform directions of motion. 
Note, however, that lane formation is not a trivial effect, 
but eventually arises only due to the smaller relative velocity
and interaction rate that pedestrians with the same walking
direction have (see Section~\ref{maceq}). In more detail,
the mechanism of lane formation can be understood as follows: Pedestrians
moving in a mixed crowd or moving against the stream will have frequent
and strong interactions.  In each interaction, the encountering
pedestrians move a little aside in order to pass each other.  This
sidewards movement tends to separate oppositely moving pedestrians.
Moreover, once the pedestrians move in uniform lanes, they will
have very rare and weak interactions.  Hence, the tendency to break up
existing lanes is negligible. Furthermore, the most stable configuration 
corresponds to a state with a {\em minimal interaction rate.} Therefore, lane
formation and minimal interaction rate are two sides of the same medal.
Nevertheless, lane formation does not occur in all driven repulsive systems.
There are certain conditions for it, which we will work out 
later on (see Section~\ref{orgi}).

\section{The Macroscopic Equation}\label{maceq}

To give an analytical description of the macroscopic dynamics of the
considered system, we will set up continuum equations
for the pedestrian densities. By indexes $a$ and $b$, 
we will distinguish different (sub-)populations 
defined by the different desired walking directions.
For the mathematical description of lane formation, it is
sufficient to focus on the one-dimensional dynamics perpendicular to the
desired walking directions.
(Imagine a projection of pedestrian dynamics on
a cross section of the walkway). 
The distribution of the $N_a$ pedestrians of population
$a$ over the locations ${x}$ of this one-dimensional 
space will be represented by the densities $\rho_a({x},t)\ge 0$.
\par
Assuming conservation of the number
\begin{equation}
  N_a = \int\limits_0^I {d}x \, \rho_a({x},t)
\label{conserv}
\end{equation}
of pedestrians in each population $a$, where $I$ denotes the spatial
extension of the system, we obtain the so-called
continuity equations\cite{Keizer}
\begin{equation}
 \frac{\partial \rho_a({x},t)}{\partial t}
 +  \frac{\partial}{\partial x} \Big[ \rho_a({x},t) {V}_{\!a}({x},t)
 \Big] = 0 \, .
\label{cont}
\end{equation}
Here, ${V}_{\!a}({x},t)$ is the average velocity of pedestrians of
population $a$ {\em perpendicular to their desired walking direction.}
In the following, we will give a rough estimate of this velocity:
It will be proportional to the frequency $\nu_a$ of interactions that
a pedestrian of population $a$ encounters with other pedestrians.
Also, it will be proportional to the average amount $\Delta x$ that
a pedestrian moves aside when evading another pedestrian. Finally,
it will be proportional to the difference of the probabilities 
$p_+$ and $p_-$ to move in positive or negative $x$-direction, respectively.
In summary, we have the relation
\begin{equation}
 V_{\!a}({x},t) = c \, \nu_a \, \Delta x \, (p_+ - p_-) \, .
\label{eins}
\end{equation}
With a prefactor $c\ne 1$ like $c(x,t)=[1-\sum_a \rho_a(x,t)/\rho_{\rm max}]$,
one can take into account that the motion is slowed down in crowded areas.
It also limits the local density to the maximum density $\rho_{\rm max}$.
\par
The interaction rate of pedestrians belonging to (sub-)population $a$ with
others is 
\begin{equation}
 \nu_a =  C_{aa} \rho_a + C_{ab} \rho_b  
\quad \mbox{with} \quad b\ne a \, ,
\label{zwei}
\end{equation}
where $C_{ba} > C_{aa}$ because of the higher 
average relative velocity
between oppositely moving pedestrians. (Although this inequality is
enough to know for the following discussion, we mention that
$C_{aa} = (D/I) \chi \sqrt{\pi \theta_a}$ and 
$C_{ab} = C_{ba} \approx (D/I) \chi (v_a + v_b)$ \cite{complex}. 
Herein, $D$ is the so-called ``total cross section'', which
corresponds to the effective diameter of a pedestrian. The factor 
$\chi$ reflects the increase of the interaction rate with
growing density due to the finite space requirements of the pedestrians.
An approximate formula is 
$\chi = 1/[1-\sum_a \rho_a /\rho_{\rm max}]^\kappa$,
where $\kappa \ge 1$ is a suitable constant and 
$\rho_{\rm max}$ is the maximum pedestrian density. Furthermore,
$\theta_a$ is the velocity variance of pedestrians belonging to
(sub-)population $a$. Finally, $v_a > \sqrt{\theta_a}$ 
and $v_b > \sqrt{\theta_b}$ represent the average velocities of
subpopulations $a$ and $b$ {\em in their desired walking directions.})
\par
We will assume that the probability of moving by $\Delta x$ in 
positive (or negative)
$x$-direction, when evading a pedestrian, is inversely proportional
to the interaction rate at position $x_+=x+\Delta x$
(or $x_-=x-\Delta x$, respectively):
\begin{equation}
 p_+ = \frac{1/\nu_a(x_+)}{1/\nu_a(x_+) + 1/\nu_a(x_-)} \, , \qquad
 p_- = \frac{1/\nu_a(x_-)}{1/\nu_a(x_+) + 1/\nu_a(x_-)} \, .
\end{equation}
A first order Taylor expansion of the nominator and the denominator
gives the following approximate relation for the difference of these
probabilities:
\begin{equation}
 (p_+ - p_-) \approx - \frac{\Delta x}{\nu_a(x)} \, 
 \frac{\partial \nu_a(x)}{\partial x}
\label{linapp}
\end{equation}
(which, strictly speaking, is restricted to cases of small gradients
$\partial \nu_a/\partial x$ as 
in the linear regime around the homogeneous solution).
Hence, with (\ref{eins}) and (\ref{zwei}), we finally obtain the following
formula for the average velocity of motion perpendicular to the
desired walking direction:
\begin{equation}
 V_{\!a}({x},t) \approx - c \, (\Delta x)^2 \bigg( C_{aa}
 \frac{\partial \rho_a}{\partial x} + C_{ab} \frac{\partial \rho_b}{\partial
   x} \bigg) \, . 
\end{equation}
Defining $S_{ab} = -(\Delta x)^2 C_{ab}$ and generalizing to an arbitrary
number $A$ of (sub-) populations gives
\begin{equation}
 V_{\!a}({x},t) \approx c \, \sum_{b=1}^A S_{ab}
 \frac{\partial \rho_b}{\partial x} \, .
\end{equation}
We may rewrite this in terms of a gradient 
\begin{equation}
 {V}_{\!a}({x},t) = c \,
 \frac{\partial S_a({x},t)}{\partial x} 
\label{gradient}
\end{equation}
of the linear density-dependent function
\begin{equation}
 S_a({x},t) = S_a^0 + \sum_b S_{ab} \, \rho_b({x},t) \, ,
\label{success}
\end{equation}
where the constants $S_a^0$ do not matter at all. Formula (\ref{success}) 
can be interpreted as a linear approximation of a more general function 
$S_a(x,t)$ of the densities.

Notice that, for higher-dimensional spaces, relation 
(\ref{gradient}) becomes a {\em potential condition}. Later on, we will see
that this is one of the conditions which must be fulfilled for optimal
self-organization. Since it is not satisfied for pedestrian crowds 
with {\em four} different desired walking directions (at intersections),
we can now understand why, in this case, there exist no optimal 
self-organized patterns of motion, which are stable. Nevertheless, collective
patterns of motion like ``rotary traffic'' can form 
temporarily \cite{book}.

\section{Self-Optimization} \label{optimi}

In the following, we will prove that, under certain conditions, the function 
\begin{equation}
 S(t) = \sum_a \int {d} x \, \rho_a({x},t) S_a({x},t)
\label{define}
\end{equation}
is a Lyapunov function which monotonically 
increases in the course of time.
Notice that $S(t)$ can be viewed as being analogous
to a thermodynamic non-equilibrium potential \cite{Graham}, 
allowing the determination of the characteristic quantities $S_{ab}$
by functional derivatives. For example, if $S_{ba} = S_{ab}$, we have 
\begin{equation}
 S_{ab} = \frac{1}{2I} \, \frac{\delta^2 S}{\delta \rho_a \delta \rho_b}
 \, .
\end{equation}

By deriving (\ref{define}) with respect to $t$,
using (\ref{success}), and properly interchanging indices $a$ and $b$,
one can eventually obtain
\begin{equation}
\frac{d S(t)}{d t}
 = \sum_{a} \int {d} x \, \frac{\partial \rho_a({x},t)}{\partial t}
\sum_b (S_{ab} +  S_{ba}) \rho_b({x},t) \, .
\label{zwi}
\end{equation}
If $S_{ab}$ is antisymmetric (i.e. $S_{ba} = -S_{ab}$), $S$ is
obviously an invariant of motion. However, in the following we will focus on
the case $S_{ba} = S_{ab}$ of symmetric interactions, which applies
to our pedestrian, granular, and trail formation examples.
Inserting (\ref{cont}) and (\ref{gradient}) into (\ref{zwi}), and
applying (\ref{success}), we get
\begin{equation}
 \frac{d S(t)}{dt} =  -2 \sum_a \int {d} x \, 
 \big[ S_a({x},t) - S_a^0 \big] \, 
 \frac{\partial}{\partial x} \bigg[ \rho_a({x},t) 
 \,c\, \frac{\partial S_a({x},t)}{\partial x} \bigg] \, .
\end{equation}
Making use of partial integration 
(for spatially periodic systems), we finally arrive at
\begin{equation}
 \frac{dS(t)}{dt} =
 2 \sum_a \int {d} x \; c \, \rho_a({x},t) \left[ 
 \frac{\partial S_a({x},t)}{\partial x} \right]^2 \ge  0 \, .
\label{endresult}
\end{equation}
This result 
establishes {\em self-optimization} for symmetical interactions
and can be easily transferred to discrete spaces 
(see Section~\ref{orgi} and Fig.~\ref{fig3}) and to higher-dimensional spaces. 
In case of slightly asymmetric interactions,
small non-linear contributions to (\ref{success}), or small diffusion,
relation (\ref{endresult}) will still be a good approximation, i.e. the
system will behave close to optimal. This is exemplified by heterogeneous
freeway traffic \cite{solid}. Later on, we will see that more or less symmetric
interactions are very natural for the kind of self-organizing systems we are
considering, here (see Section~\ref{results}).
\par
Notice that (\ref{endresult}) looks similar to dissipation functions in
thermodynamics \cite{Keizer,Rayleigh,Onsager,deGroot,Graham}. 
If we interpret the function $dS(t)/dt$
as a measure of dissipation per unit time in the system,
equation~(\ref{endresult}) immediately implies that {\em the system approaches
a stationary state of minimal dissipation,} since $S(t)$ is bounded for any
finite system (which means $dS(t)/dt = 0$ in the limit of large 
times $t$). 
This may be viewed as a generalization of the related Onsager principle
of minimal dissipation of entropy \cite{Onsager,deGroot}, 
dating back to Lord Rayleigh \cite{Rayleigh}. In
contrast to the non-linear, far-from equilibrium systems considered above,
the Onsager principle applies to linearly treatable, close-to-equilibrium
systems only, which usually tend to approach a homogeneous state.
\par
According to (\ref{endresult}), the stationary solution
$\rho_a^{\rm st}({x})$ is characterized by
\begin{equation}
 c \, \rho_a^{\rm st}({x}) = 0 \qquad \mbox{or} \qquad 
 \frac{\partial S_a^{\rm st}({x})}{\partial x} = \sum_b S_{ab} 
 \frac{\partial \rho_b^{\rm st}({x})}{\partial x} = {0}
\label{statio}
\end{equation}
for all $a$, which is fulfilled by homogeneous or by
step-wise constant solutions. Hence, {\em the stationary solution can
be non-homogeneous, but nevertheless satisfies 
minimal dissipation,} which is quite interesting.
We would not have recognized this, if we would have derived
the relations (\ref{statio}) for the stationary solution directly from
the continuity equation (\ref{cont}) 
with (\ref{gradient}) and (\ref{success}). 

\section{The Relation with Game Theory} \label{gameth}

Physicists have recently gained a considerable interest in applications
of methods from statistical physics and nonlinear dynamics to 
biological \cite{eshel,biol1a,biol1b,biol1c,biol2,biol3,biol4,biol5,biol6,%
Drasdo,ants1,ants2}, 
economic \cite{mandel,Bouchaud,solomon,stanley,peinke,zhang,Stauf,We91}, 
and social \cite{We91,Hub,Hel93,Galam,Ebeling,gamedyn,quantsoc,frank,gd} 
systems. Hence, it is worth stressing that the above model can be applied to 
such kinds of systems, if we give a more general interpretation to it. 
First of all, instead of particles or pedestrians, we may have other kinds
of entities, which we again need to subdivide into uniform (sub-)populations,
according to their behaviors. Second, the space may be rather abstract than
real, for example, a behavioral space or an opinion spectrum 
\cite{quantsoc,Hel93}.
The same applies to motion, which may correspond to a change of behavior or
opinion. (In the case of trail formation, a point $\vec{x}$ in space
even corresponds to a {\em path} connecting two places, namely
a pedestrian source with one of their destinations.)
 
Having again a look at the relation (\ref{gradient}), 
it makes sense to interpret
the function $S_a({x},t)$ as the ``(expected) success''
per unit time for an entity of population $a$ at location ${x}$,
since it is plausible that the entities move into the direction of the
greatest increase of success (which is the direction of the gradient).
Furthermore, one can give a more concrete meaning to the coefficients $S_{ab}$:
If an entity of kind $a$ interacts with
entities of kind $b$ at a rate $\nu_{ab}$ and the associated
outcome of the interaction can be quantified by some ``payoff'' $P_{ab}$, we
have the relation $S_{ab} = \nu_{ab}P_{ab}$.
Positive payoffs $P_{ab}$ belong to attractive or, more general, {\em profitable}
interactions between populations $a$ and $b$, while negative ones 
correspond to repulsive or {\em competitive} interactions. We think, it is quite
surprising that, based on competitive interactions, there can be 
self-organized and even optimal system states at all 
(see Section~\ref{results}).

Notice that, despite of the mentioned relations with game theory,
our model differs from the conventional
game dynamical equations \cite{Ebeling,gamedyn,quantsoc} 
in several respects:
\begin{enumerate}
\item We have a topology (like in the {\em game of life} \cite{life,may}),
but define abstract games for interactive
motion in space with the possibility of local agglomeration
at a fixed number of entities in each population.
\item The payoff does not depend on the variable that the individual entities 
can change (i.e. the spatial coordinate $x$).
\item Individuals can only improve their success by redistributing
themselves in space.
\item The increase of success is not proportional to the
difference with respect to the global average of success, 
but to the local change of success in a population.
\end{enumerate}
Moreover, below we will establish a new connection between self-optimization 
and self-organization (see Section~\ref{results}), 
which we consider to be important.

\section{Self-Organization} \label{orgi}

Our generalized model allows to describe 
all kinds of different combinations between attractive or
profitable and repulsive or competitive interactions within and among the
different populations. It is, therefore, desireable to know the exact
conditions under which the corresponding 
system forms a self-organized, i.e. a non-homogeneous
state. In order to derive these, we will carry out a linear stability analysis
around the homogeneous stationary solution $\rho_a^{\rm hom}= N_a/I$,
where $I$ again denotes the spatial extension of the system.
For simplicity, we will restrict
ourselves to the case of two (sub-)populations $a, b \in \{1,2\}$.

We start with the continuity equation
\begin{equation}
 \frac{\partial \rho_a}{\partial t}
 + \frac{\partial}{\partial x} \bigg[ \rho_a c \,
 \frac{\partial S_a}{\partial x} \bigg] = c \, D_a 
 \frac{\partial^2 \rho_a}{\partial x^2} \, ,
\label{contdiff}
\end{equation}
where we have introduced an additional 
diffusion term with diffusion coefficient $c\,D_a$
on the right-hand side, since we will discuss the effect of fluctuations
later on. Next, we write down the corresponding linearized partial
differential equations:
\begin{eqnarray}
 \frac{\partial \rho_a}{\partial t} 
 &=& - c\,\rho_a \frac{\partial^2 S_a}{\partial x^2}
 + c \, D_a \frac{\partial^2 \rho_a}{\partial x^2} \nonumber \\
 &=& - \sum_{b=1}^2 c\, \rho_a S_{ab} \frac{\partial^2 \rho_b}{\partial x^2}
 + c\,D_a \frac{\partial^2 \rho_a}{\partial x^2} \, .
\end{eqnarray} 
Inserting into these equations the {\em ansatz} 
\begin{equation}
 \rho_a(x,t) = \rho_a^{\rm hom} + \tilde{\rho}_a \, 
 \mbox{e}^{{\rm i}kx + \lambda t} \, ,
\end{equation}
where $k$ has the meaning of a wave number,
leads to the following linear eigenvalue problem with
eigenvalue $\lambda$:
\begin{equation}
\lambda \left( 
\begin{array}{c}
\tilde{\rho}_1 \\
\tilde{\rho}_2 
\end{array} \right) =
\left( \begin{array}{cc}
A_{11} & A_{12} \\
A_{21} & A_{22}
\end{array} \right)
 \left( 
\begin{array}{c}
\tilde{\rho}_1 \\
\tilde{\rho}_2 
\end{array} \right) \, .
\end{equation}
Herein, we have $A_{11} = c\, k^2 (\rho_1^{\rm hom} S_{11} - D_1)$, 
$A_{12} = c\, k^2 \rho_1^{\rm hom} S_{12}$, 
$A_{21} = c\, k^2 \rho_2^{\rm hom} S_{21}$, and
$A_{22} = c\, k^2 (\rho_2^{\rm hom} S_{22} - D_2)$.
The linear system of equations can be solved for the two eigenvalues
\begin{equation}
 \lambda_{1/2} = \frac{A_{11}+A_{22}}{2} \pm \frac{\sqrt{(A_{11}+A_{22})^2 
 - 4 (A_{11}A_{22} - A_{12}A_{21}) }}{2} \, .
\end{equation}
In order for the homogeneous solution to be stable, the real values of
both eigenvalues need to be negative. This requires
\begin{equation}
(A_{11}+A_{22}) < 0 \quad \mbox{and} \quad
(A_{11}A_{22} - A_{12}A_{21}) > 0\, .
\label{relations}
\end{equation}

Notice that, in the case of an unstable eigenvalue, it is the mode with
the largest wave number (i.e. with the shortest wave length) that grows
fastest. This somewhat unrealistic behavior is a consequence of 
simplifications made, namely the linear approximation underlying 
relation (\ref{linapp}).
In reality, the spatial extension of the entities will
introduce a natural cutoff for the wave lengths. One may consider this in
equation (\ref{contdiff}) by additional spatial derivatives of higher
(e.g. fourth) order. However, one can easily
circumvent these problems by setting up a discrete version of the model
(where the spatial discretization should be chosen in accordance with
$\Delta x$). 

The results of Figures~\ref{fig3} and \ref{fig4}, 
for example, were obtained with the following discrete analogue
of the model defined by Eqs. (\ref{cont}), (\ref{gradient}), 
and (\ref{success}).
We assumed a periodic lattice
with $I$ lattice sites $x\in\{1,\dots,I\}$ and two
populations $a\in\{1,2\}$ with a total of $N=N_1+ N_2 \gg I$ entities.
(In the figures we used $I=40$ and $N_1 = N_2 = 200$.) 
At time $t=0$,
our simulations started with a random initial distribution of the entities.
Then, we repeatedly applied the following update steps to determine the
distribution of entities at time $t+1$: 
\begin{enumerate}
\item Calculate the successes 
\begin{equation}
 S_a(x,t) = S_a^0 + \sum_b S_{ab} \, \frac{n_x^b(t)}{I} \, ,
\end{equation}
where $n_x^b(t) = \rho_b(x,t) I$ represents the
number of entities of population $b$ at site $x$.
\item For each entity $\alpha$,
determine a random number $\eta_\alpha$ that is uniformly
distributed in the interval $[0,S_{\rm max}]$ with a large constant 
$S_{\rm max}$ ($S_{\rm max}=20$ in the figures).
\item Move entity $\alpha$ belonging to population $a$ from site $x$ to site
$x+1$, if 
\begin{equation}
 c(x+1,t)[S_a(x+1,t) - S_a(x-1,t)] > \eta_\alpha(t) \, ,
\label{label1}
\end{equation}
but to site $x-1$, if 
\begin{equation}
 c(x-1,t)[S_a(x-1,t) - S_a(x+1,t)] > \eta_\alpha(t) \, .
\label{label2}
\end{equation} 
(Figure~\ref{fig3} and \ref{fig4} are for
$c(x,t)=1$.) In cases, where we assumed errors in the estimation of the
expected success, we replaced $S_a(x,t)$ by $S_a(x,t) + \xi_\alpha(t)$,
where $\xi_\alpha(t)$ is determined according to some
probablility distribution. 
\end{enumerate}
We applied a random sequential update rule, 
which is most reasonable \cite{Bernardo}.
However, a parallel update, which defines a simple cellular automaton 
\cite{Wolfram,Stauffer}, yields 
qualitatively the same results (even nicer looking ones, because it does not
have the fluctuations caused by a random update).

\section{Results and Discussion} \label{results}

Let us first discuss the case $D_a \approx 0$ of negligible diffusion.
For this case, the relations (\ref{relations}) imply
that the homogeneous solution of the model
(cf. Figure~\ref{fig3}a) is unstable if
\begin{equation}
 \rho_1^{\rm hom} S_{11} + \rho_2^{\rm hom} S_{22} > 0 
\label{cond1}
\end{equation}
or
\begin{equation} 
 S_{12} S_{21} > S_{11} S_{22}  \, .
\label{cond2}
\end{equation}
In other words, if one of the conditions (\ref{cond1}) or (\ref{cond2})
is fulfilled, the stable stationary solution is a self-organized,
non-homogeneous state which, according to (\ref{statio}), 
corresponds to complete segregation (or aggregation). 
If the populations interact
in a symmetric way, the underlying self-organization process is related to 
self-optimization (see (\ref{endresult})). 
Condition (\ref{cond1}) is satiesfied by systems in which the interactions
within the (sub-) populations are attractive or profitable. Such systems show
always some form of agglomeration (see Figs.~\ref{fig3}c and d).

In the following, we focus on the more common and much more interesting cases 
where (\ref{cond1}) is not valid. This will allow us to answer the question, 
why self-organizing systems of the considered type tend to reach an
optimal state. The solution is: ``because they tend to be symmetric!''. 
This can be seen as follows: Introducing 
\begin{equation}
 \overline{S} = \frac{S_{12} + S_{21}}{2} \quad \mbox{and} \quad
 \Delta S = \frac{S_{12}-S_{21}}{2} \, ,
\end{equation} 
we have 
\begin{equation}
S_{12} = \overline{S} +  \Delta S 
\quad \mbox{and} \quad
S_{21} = \overline{S} - \Delta S \, , 
\end{equation}
and condition (\ref{cond2}) becomes 
\begin{equation}
 (\overline{S}+\Delta S)(\overline{S}-\Delta S) 
 = \overline{S}^{\,2} - (\Delta S)^2 > S_{11}S_{22} \, .
\end{equation}
Since the interaction strengths 
$|S_{11}|$, $|S_{22}|$, $|S_{12}|$, $|S_{21}|$, and 
$|\overline{S}|$ will normally have the same order of magnitude, this
condition for self-organization can only be fulfilled for small $|\Delta S|$,
i.e. 
\begin{equation}
 S_{12} \approx S_{21} \, . 
\end{equation}
Hence, there is a tight connection between self-organization and
self-optimization
in the considered kinds of systems: {\em If there is 
self-organization, 
it is likely to come with optimality,} at least approximately. 
For this reason, 
one may speak of a ``self-organized system with optimality'' or more compact of
``self-organized optimality''
(although there is no immediate relation
with ``self-organized criticality''\cite{SOC}). However, the more
precise term is probably ``optimal 
self-organi\-za\-tion''.\footnote{This implies that there are also
``non-optimal'' forms of self-organization like in systems for which
no Lyapunov function exists. A typical example are systems with
oscillating or chaotic states.}
A good example is uni-directional multi-lane traffic
of cars and trucks \cite{solid}, which develops a self-organized, 
coherent state only in a small density range, which is also
characterized by minimal interactions 
(minimal lane-changing and interaction rates). From the above, we conclude
that, only in this density range, the interactions of cars and trucks
become sufficiently symmetric to give rise to self-organization and
self-optimization. This conclusion is quite reasonable, since, according to
the assumed traffic model, the interactions of
cars and trucks are more symmetric when their average velocities 
are similar.
\par
In our pedestrian example, we find optimal self-organization, 
since the symmetry condition is satisfied exactly,
and condition (\ref{cond2}) is also
fulfilled (see Eq. (\ref{zwei}) and below).
Note, however, that there are many other examples of segregation in
the natural and social sciences \cite{granular,seg1,seg2,seg3,We91}.
We point out that the spatial regions occupied by one population need 
not be connected (cf. Figures~\ref{fig3}b to \ref{fig3}d), and that 
the corresponding configuration may correspond to
a {\em relative} optimum as in Figures~\ref{fig3}c and \ref{fig3}d.
If $S_{aa}< 0$ for all $a$, the distributions $\rho_a^{\rm st}({x})$ 
tend to be flat, as in the case of 
lane formation by repulsive pedestrian interactions (Figure~\ref{fig3}b).
Instead, we have agglomeration (local clustering), 
if $S_{aa} >0$ for all $a$ (Figures~\ref{fig3}c and \ref{fig3}d).
Figure \ref{fig3}c describes the
segregation of populations with repulsive 
interactions (e.g. ``ghetto formation''). 
The case of Figure~\ref{fig3}d allows to understand the conjectured
optimality of trail systems which, based on attractive interactions,
result by a bundling of
trails ending at different destinations \cite{trail}.
\par
Finally, we discuss the influence of fluctuations. 
Including the effect of diffusion,
the conditions for self-organization will become
\begin{equation}
 \rho_a^{\rm hom} S_{aa} + \rho_b^{\rm hom} S_{bb} > D_a + D_b
\label{cond3}
\end{equation} 
or
\begin{equation}
 S_{ab} S_{ba}\rho_a^{\rm hom} \rho_b^{\rm hom} 
 >  (S_{aa} \rho_a^{\rm hom} - D_a)(S_{bb} \rho_b^{\rm hom} - D_b) \, .
\label{cond4}
\end{equation}
Hence, {\em large\/} diffusion coefficients will produce homogeneous
equilibrium states, and the principle (\ref{endresult}) 
of self-optimization is not valid anymore 
(see Figure~\ref{fig3}e).\footnote{For example, 
lane formation in streams of oppositely moving
particles can be suppressed by sufficiently strong fluctuations,
giving rise to ``frozen'' (blocked) states \cite{freezing}.}
However, {\em small diffusion can further self-organization.} Not only
will the system be able to escape relative optima and
eventually reach the global optimum (cf. Figures~\ref{fig3}f and
\ref{fig4}), although the
interactions are short-ranged (see Eq. (\ref{gradient})).
According to (\ref{cond4}), small diffusion can also reduce the 
stability of the homogeneous 
stationary solution, which is quite surprising.

\section{Summary and Outlook} \label{outlook}

Our investigations were motivated by the question why many self-organized
systems seem to optimize certain macrosopic quantities, for which we have
given a number of realistic examples. In order to explain this, we have
derived macroscopic equations for lane formation in systems 
of oppositely moving driven particles or pedestrians. We could show that,
in cases of repulsive interactions, the system tends to minimize
the interaction rate and the intensity of interactions. 
The applicability of the
model, however, could be extended to cases of attractive interactions and,
using game theoretical ideas, also to competitive or profitable
interactions in many kinds of non-physical systems like biological,
economic, or social ones. 

After having shown that many driven systems can be represented as
a game between interacting (sub-)populations, we have constructed a
functional for such systems, which is related to
thermodynamic non-equilibrium potentials and can
be interpreted as overall (expected) success. In cases of symmetric
interactions among the populations,
this function increases monotonically in the course of time,
meaning that the overall success of these systems is optimized. 
In other words, as individual entities are trying to maximize
their {\em own} success, these systems tend to reach a state with
the highest {\em global} success, which is not trivial at all. 
Since the form of the increasing
function reminds of a generalized thermodynamic dissipation 
function, one can also say that the system approaches a
state of minimal dissipation (which may be considered as a generalized
Onsager principle). This principle of ``minimal waste of energy''
may be particularly interesting for biology, where it can be conjectured 
that organisms use energy very efficient. For example, it is known that
pedestrians tend to move at the speed which is least energy consuming
\cite{Weidmann}. We think it is worth pointing out that
there are quite a number of living systems (for which we
have given some realistic examples) to which existing
methods and notions from thermodynamics 
can be successfully applied, if they are generalized in a suitable way. 
It would be surely interesting to look for other systems, for which similar
results or principles can be found. 

Here, we obtained that
the precondition for self-organization with self-optimization
is symmetric interactions.
However, we could give arguments indicating that {\em the considered systems,
if they self-organize at all, are also (more or less) symmetric and, hence,
behave (close to) optimal.} 
Therefore, the phenomenon of ``self-organized optimality'' or
``optimal self-organization'', as we call it, is expected
to be quite common in nature. Already for
two symmetrically interacting populations, one can classify more than 
ten different
situations (dependent on whether $S_{11}$, $S_{22}$, and $S_{12}=S_{21}$
are smaller or greater than zero, 
and whether conditions (\ref{cond1}) or (\ref{cond2})
are fulfilled or not). 
This includes quite surprising cases
as for repulsive interactions within each population
and stronger attractive interactions between them
(e.g. $S_{11}=S_{22}=-1$, $S_{12}=S_{21}=2$), which leads to agglomeration
analogous to Figure~\ref{fig3}d rather than homogeneously distributed,
mixed populations as in Figure~\ref{fig3}a. Apart from the results displayed
in Figure~\ref{fig3}, there are also cases where one population agglomerates,
but the other one is distributed homogeneously (if we choose $S_{11}$ different
from $S_{22}$). In systems with diffusion, the
variety of different cases covered by the above approach is even greater.
In particular, we have observed that, while large noise generally destroys
self-organized solutions, {\em small noise can further self-organization}
in the considered systems, which is surprising.

Finally, we point out that our results are relevant for practical applications.
For example, when optimizing multi-agent systems (like 
the coordination of vehicle or air traffic,
or the usage of CPU time in computer networks),
it is desireable to apply control strategies that
are insensitive to system failures (like the temporary breakdown of a control
center). Hence, it would be favourable if the system would
optimize its state by means of the interactions
in the system. For this, one needs to implement a suitable
type of interactions (namely, symmetric ones), which can be reached by
technical means (intelligent communication devices determining the
proper actions of the interacting entities). 
Notice that this kind of multi-agent optimization is
decentralized and, therefore, much more robust than classical,
centralized control approaches.

{\em Acknowledgments:} 

D.H. is grateful to the DFG for financial support
by a Heisenberg scholarship. 
This work was in part supported by OTKA F019299 and FKFP 0203/1997.
The authors also want to thank Ill\'{e}s J. Farkas
for producing Figure~\ref{fig2} and
Robert Axelrod, Eshel Ben-Jacob, and Martin Treiber for valuable comments.

\clearpage

\clearpage
\unitlength10mm
\abbl{\epsfig{width=8.0\unitlength, angle=90, 
bbllx=2pt, bblly=2pt, bburx=478pt, bbury=238pt, clip=,
file=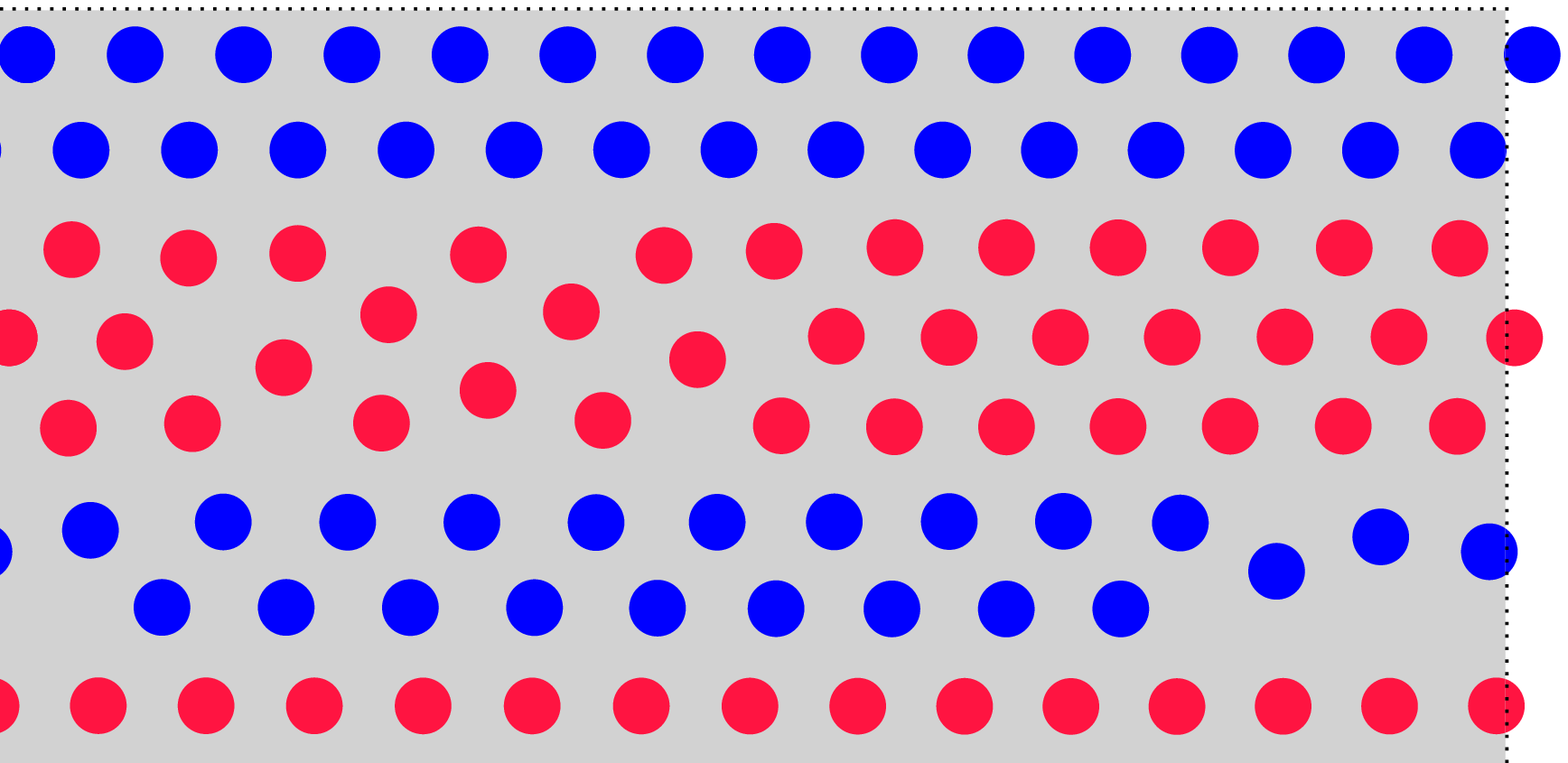}}
{}{Formation of lanes of
uniform walking directions in crowds of
oppositely moving pedestrians. Red circles represent pedestrians walking
up the street, blue ones move downwards.
The simulation assumed periodic boundary conditions,
but we could also use walls on both sides or randomly feed
pedestrians into the upper and lower boundaries of the simulation area, without
destroying the effect of lane formation. (To see this, you may check
out the Java applets supplied on this internet site:
{\tt http://www.theo2.physik.uni-stuttgart.de/ helbing.html})\label{fig2}}{}
\clearpage
\unitlength4.2mm
\clearpage
\unitlength1.6cm
\begin{figure}[htbp]
\begin{center}
\begin{picture}(5,5.6)(-0.7,-0.5)
\thicklines
\put(0,0){\circle*{0.15}}
\put(0,4){\circle*{0.15}}
\put(4,0){\circle*{0.15}}
\put(4,4){\circle*{0.15}}
\put(0,0){\line(1,1){1}}
\put(0,4){\line(1,-1){1}}
\put(4,0){\line(-1,1){1}}
\put(4,4){\line(-1,-1){1}}
\put(1,1){\line(1,0){2}}
\put(1,1){\line(0,1){2}}
\put(3,3){\line(-1,0){2}}
\put(3,3){\line(0,-1){2}}
\end{picture}
\end{center}
\caption[]{Schematic representation of a human trail system (solid lines)
evolving between four entry points and destinations (full circles)
on an initially homogeneous ground \cite{trail}.
When the frequency of trail usage
is small, the direct way system (consisting of the four
ways along the edges and the two diagonal connections) is too long to
be maintained in competition with the regeneration
of the vegetation. Here, by bundling of trails, the frequency of usage
becomes large enough to support the depicted trail system. It corresponds
to the optimal compromise between the diagonal ways and the
ways along the edges, supplying maximum walking comfort at a minimal
detour of 22\% for everyone. 
In this example, it is the discomfort of walking multiplied by the 
length of the individual ways that is minimized.
\label{tr}}
\end{figure}
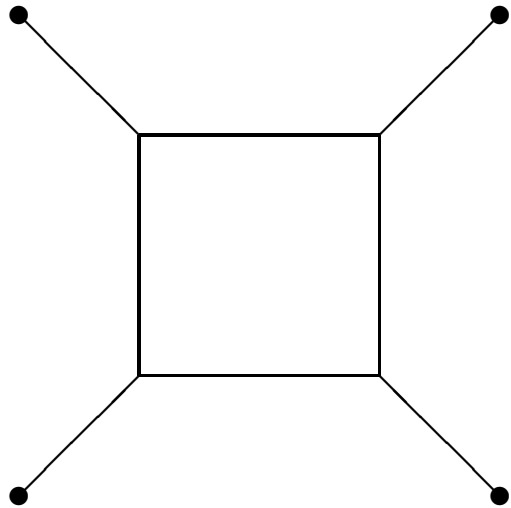
\clearpage
\unitlength10mm
\begin{center}
\begin{picture}(17,21.5)
\put(0,22.6){\epsfig{height=7.5\unitlength, angle=-90,
      file=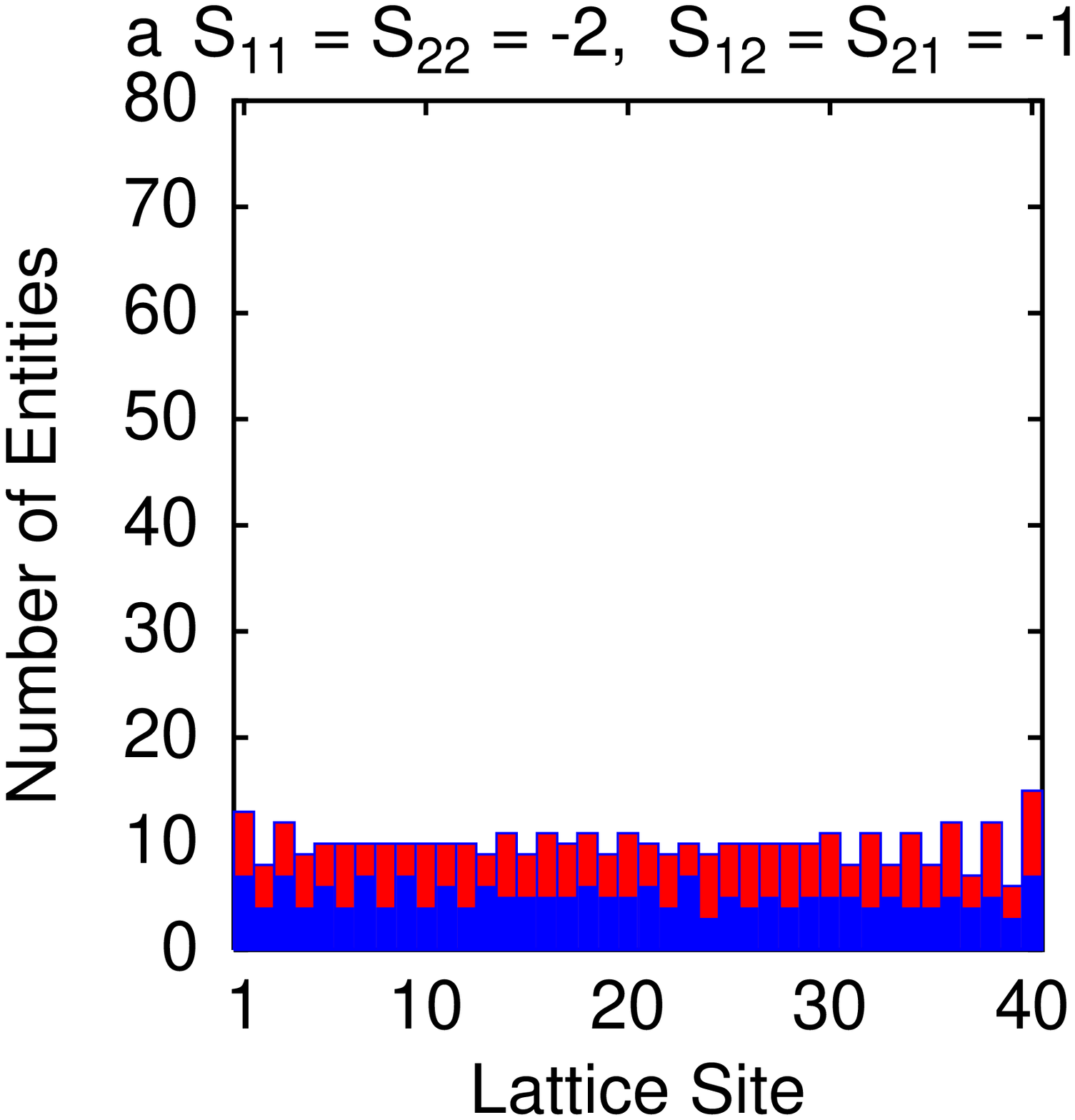}}
\put(1.8,21.9){\epsfig{height=4.8\unitlength, angle=-90,
      file=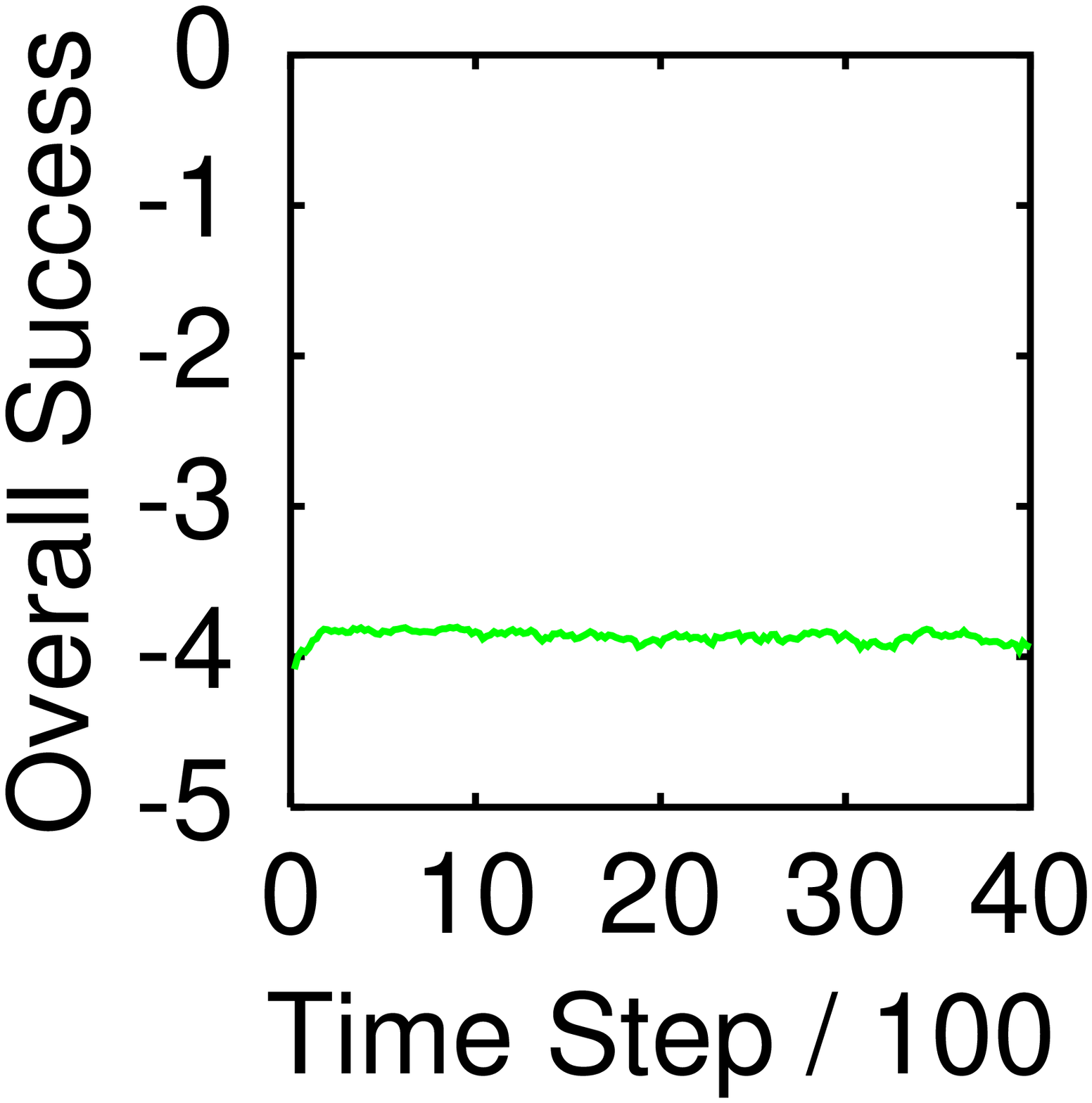}}
\put(8.0,22.6){\epsfig{height=7.5\unitlength, angle=-90,
      file=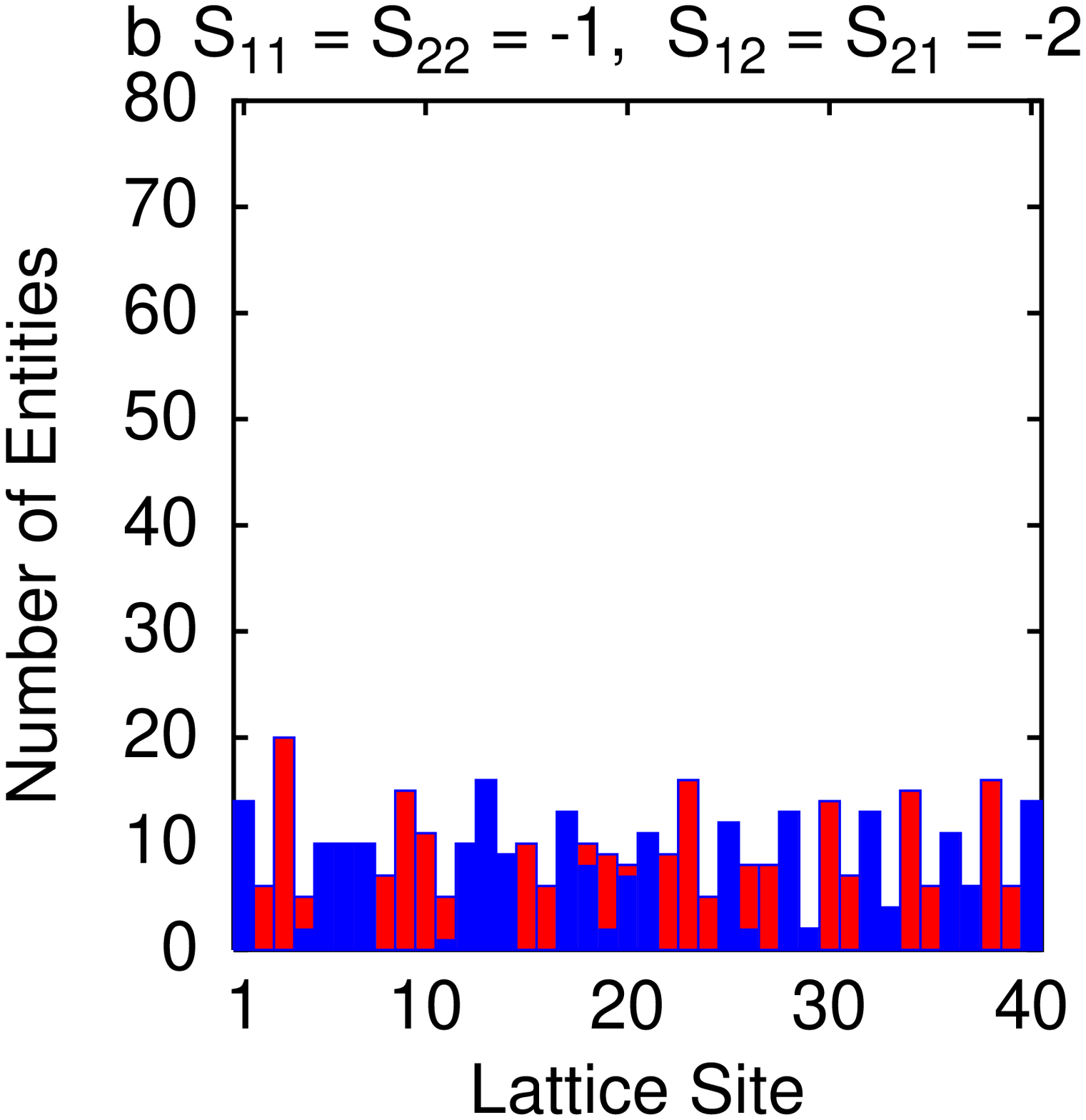}} 
\put(9.8,21.9){\epsfig{height=4.8\unitlength, angle=-90,
      file=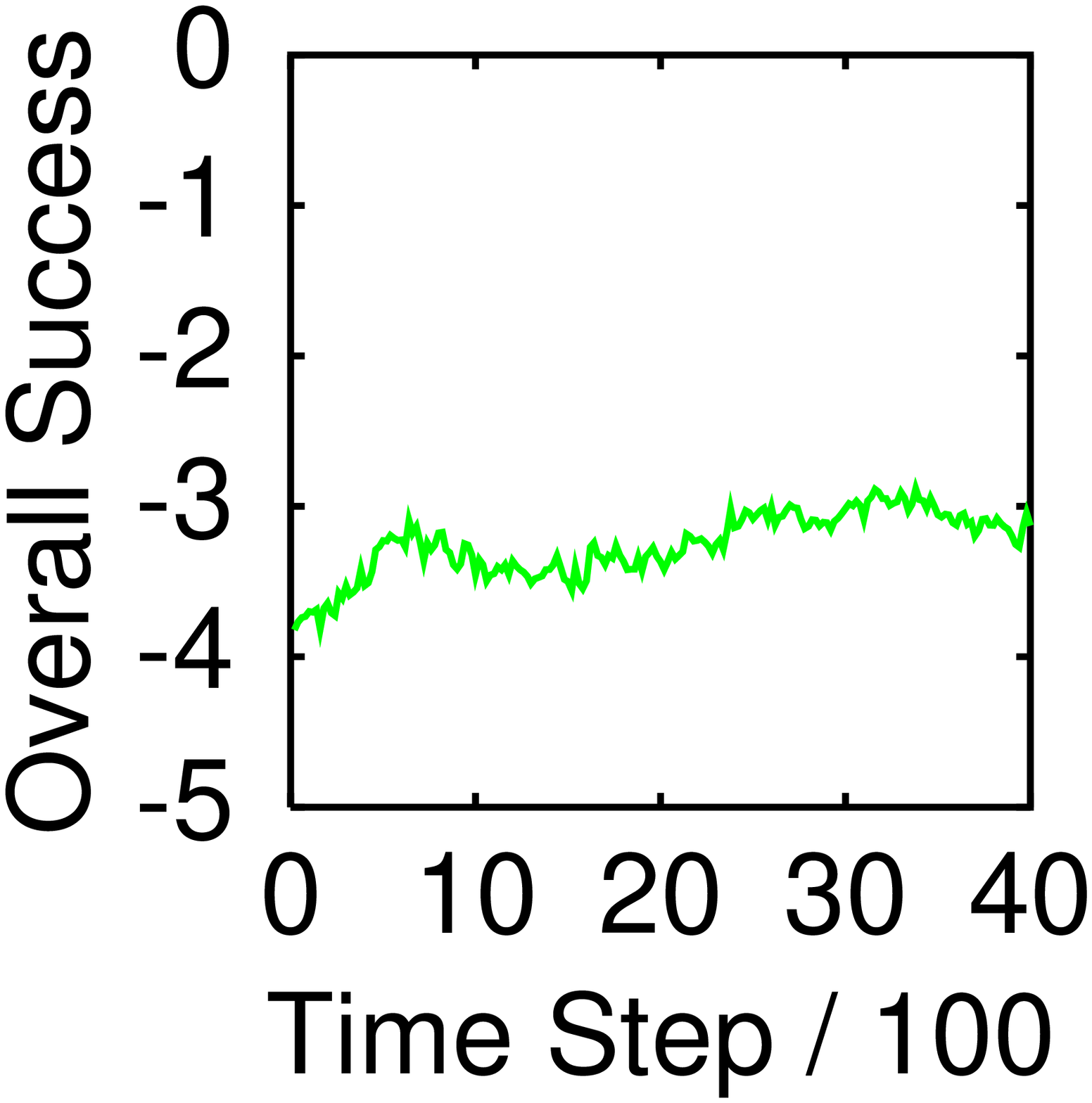}} 
\put(0,14.8){\epsfig{height=7.5\unitlength, angle=-90,
      file=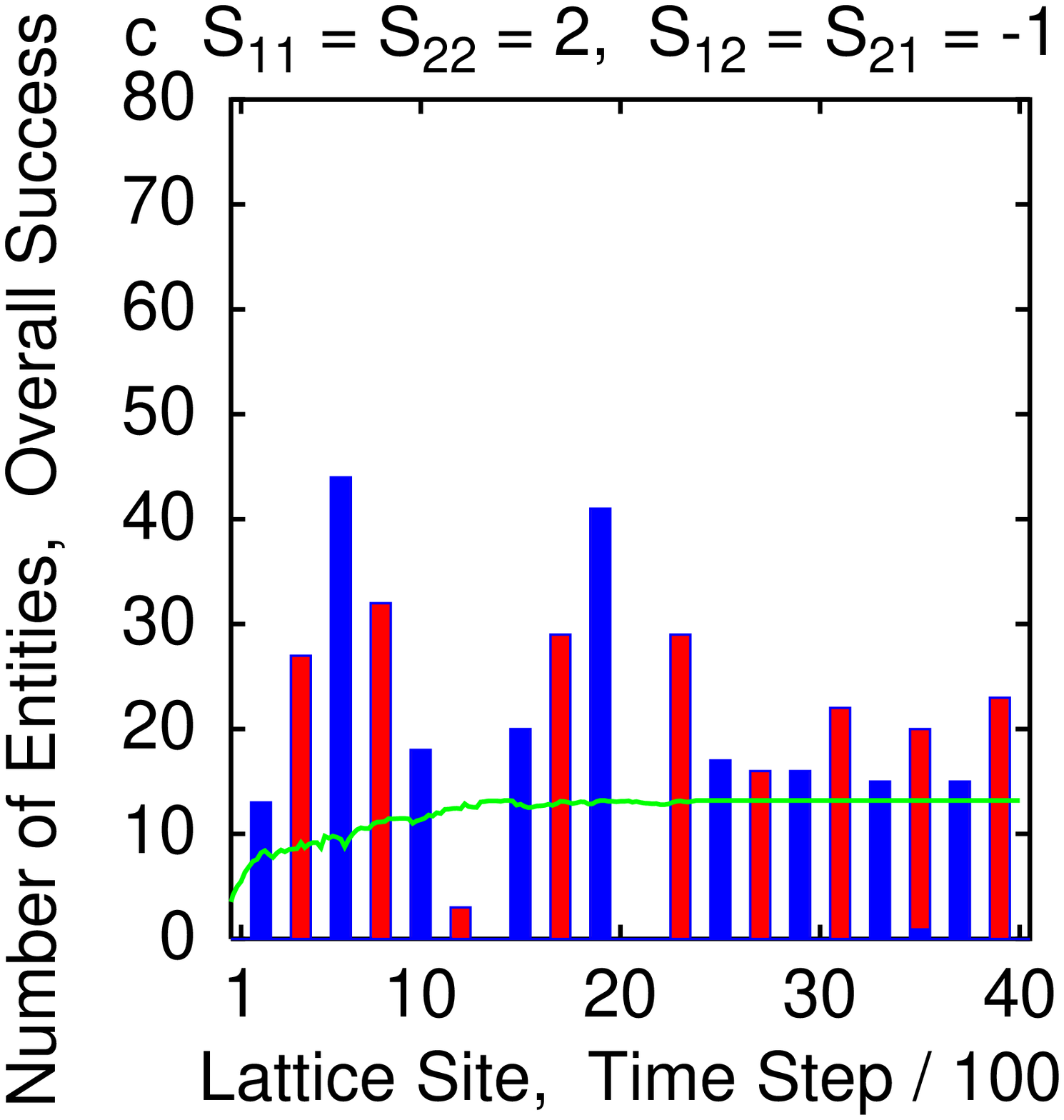}} 
\put(8.0,14.8){\epsfig{height=7.5\unitlength, angle=-90,
      file=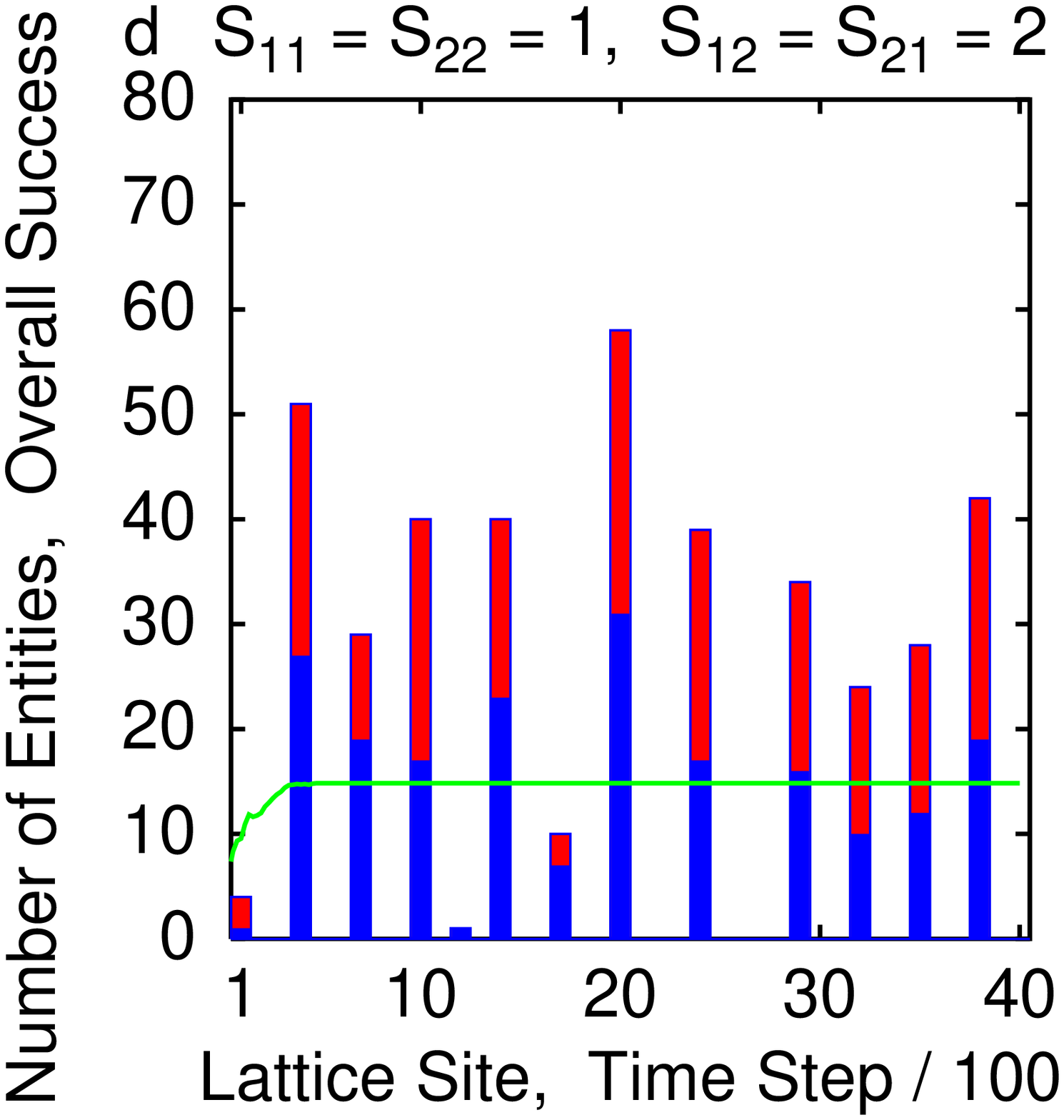}}
\put(0,7.0){\epsfig{height=7.5\unitlength, angle=-90,
      file=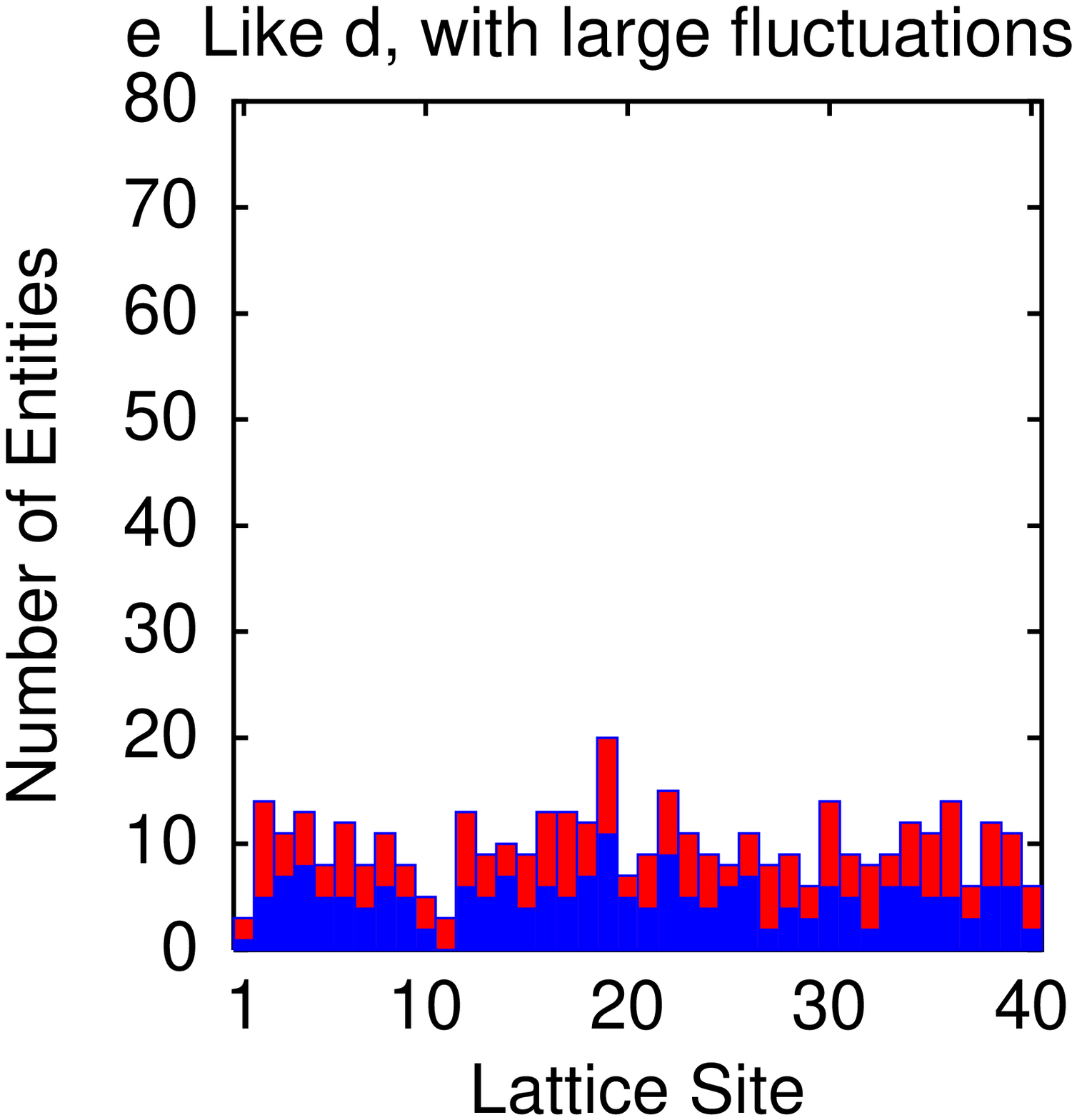}}
\put(1.8,6.3){\epsfig{height=4.8\unitlength, angle=-90,
      file=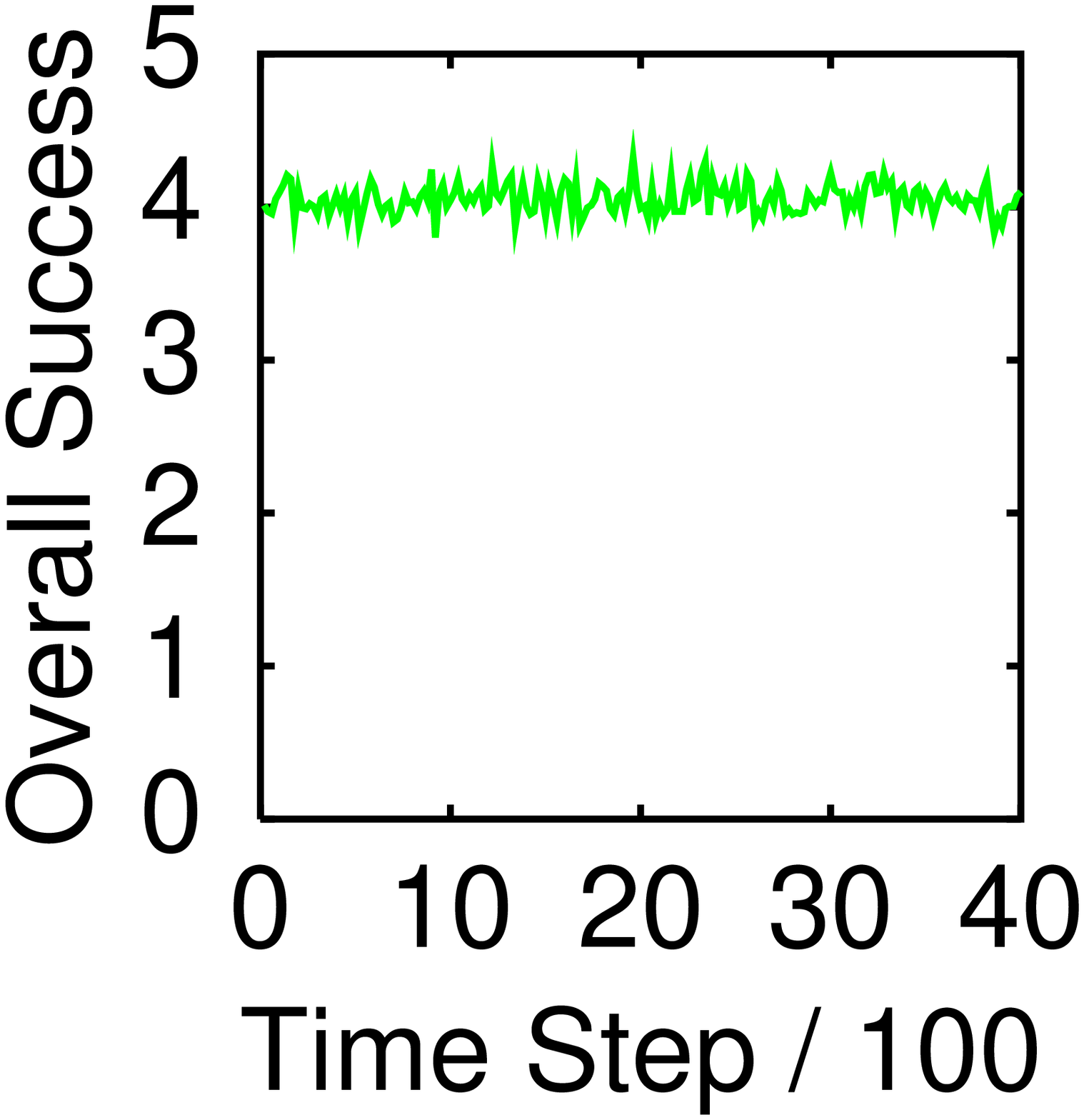}}
\put(7.7,7.0){\epsfig{height=7.85\unitlength, angle=-90,
      file=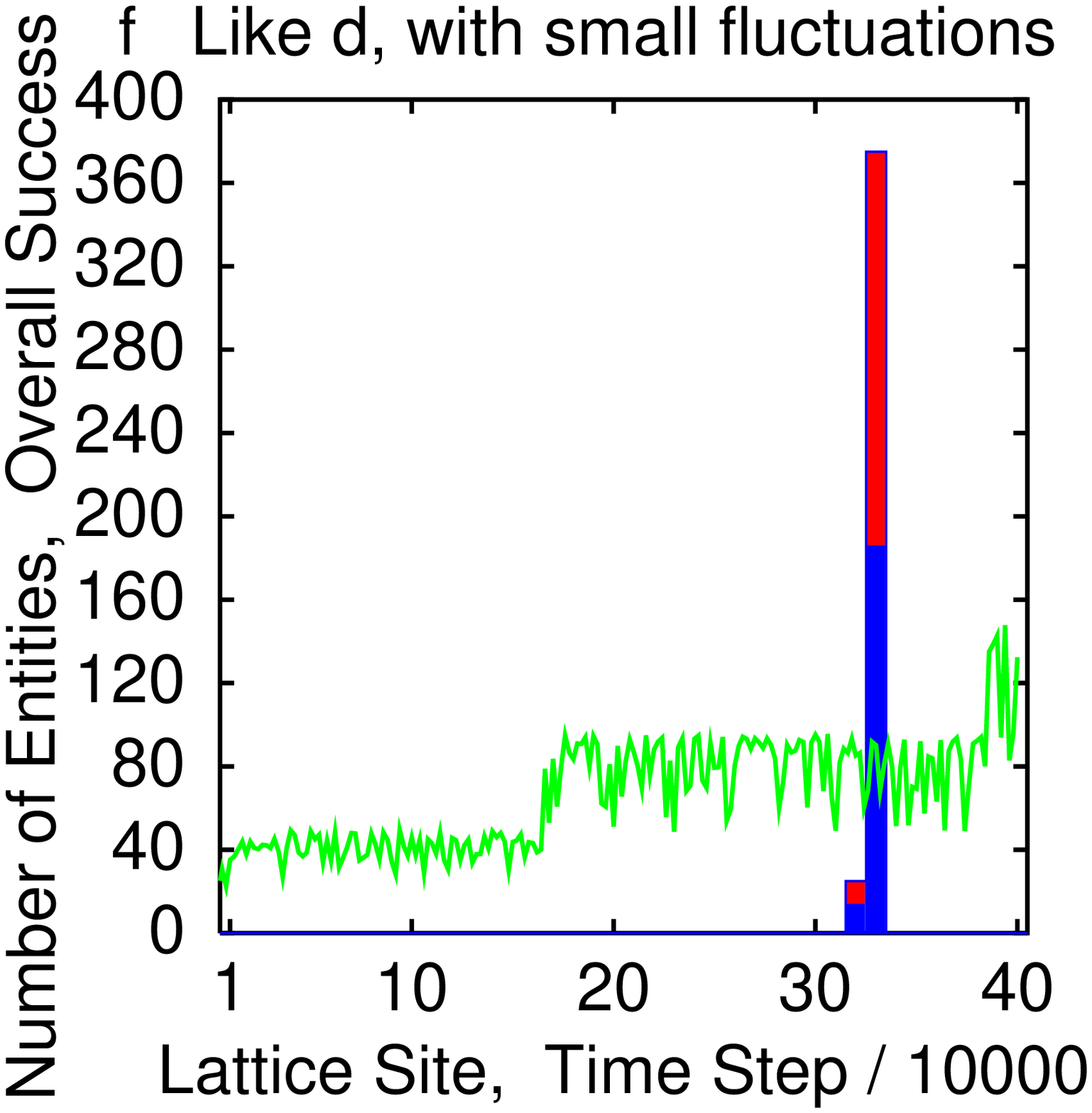}}
\end{picture}
\vspace*{-0.6\unitlength}
\end{center}
\clearpage
\begin{figure}[htbp]
\caption[]{Illustration of various forms of self-optimization
for two different populations:
{\bf a} Homogeneous distribution in space, {\bf b} segregation of 
populations without agglomeration, {\bf c} repulsive agglomeration, 
{\bf d} attractive agglomeration. Cases {\bf b} 
to {\bf d} are examples of ``optimal self-organization'' or
``self-organized optimality'', since
the finally evolving optimal states are 
self-organized, non-homogeneous states. In contrast to the results
displayed in {\bf a} to {\bf d}, in {\bf e} and {\bf f} we have additionally
introduced fluctuations corresponding
to errors in the estimation of success $S_a(x,t)$. Large fluctuations
destroy the tendency of self-optimization and produce a homogeneous
distribution of entities (see {\bf e}), whereas small fluctuations help
to escape relative (``local'') optima, 
leading to a continuation of the agglomeration process until the
{\em absolute} (``global'') optimum is reached (see Figure~\ref{fig4}).
\par
The above figures show 
the numbers $n_x^1$ and $n_x = (n_x^1+n_x^2)$ of entities as a function
of the lattice site $x$ at time $t=4000$ (in {\bf f}: $t=40000$)
and the evolution of the 
overall success $S(t)$ as a function of time $t$. The fluctuations
around the monotonic increase of $S(t)$ in {\bf a} to {\bf d} are caused by 
the fluctuations $\eta_\alpha$ (see Eqs.~(\ref{label1}), (\ref{label2}))
and the random sequential update.
\label{fig3}}
\end{figure}
\unitlength6mm
\begin{figure}[htbp]
\begin{center}
\begin{picture}(14,10)
\put(-2,13.5){\epsfig{width=14.0\unitlength, angle=-90,
      file=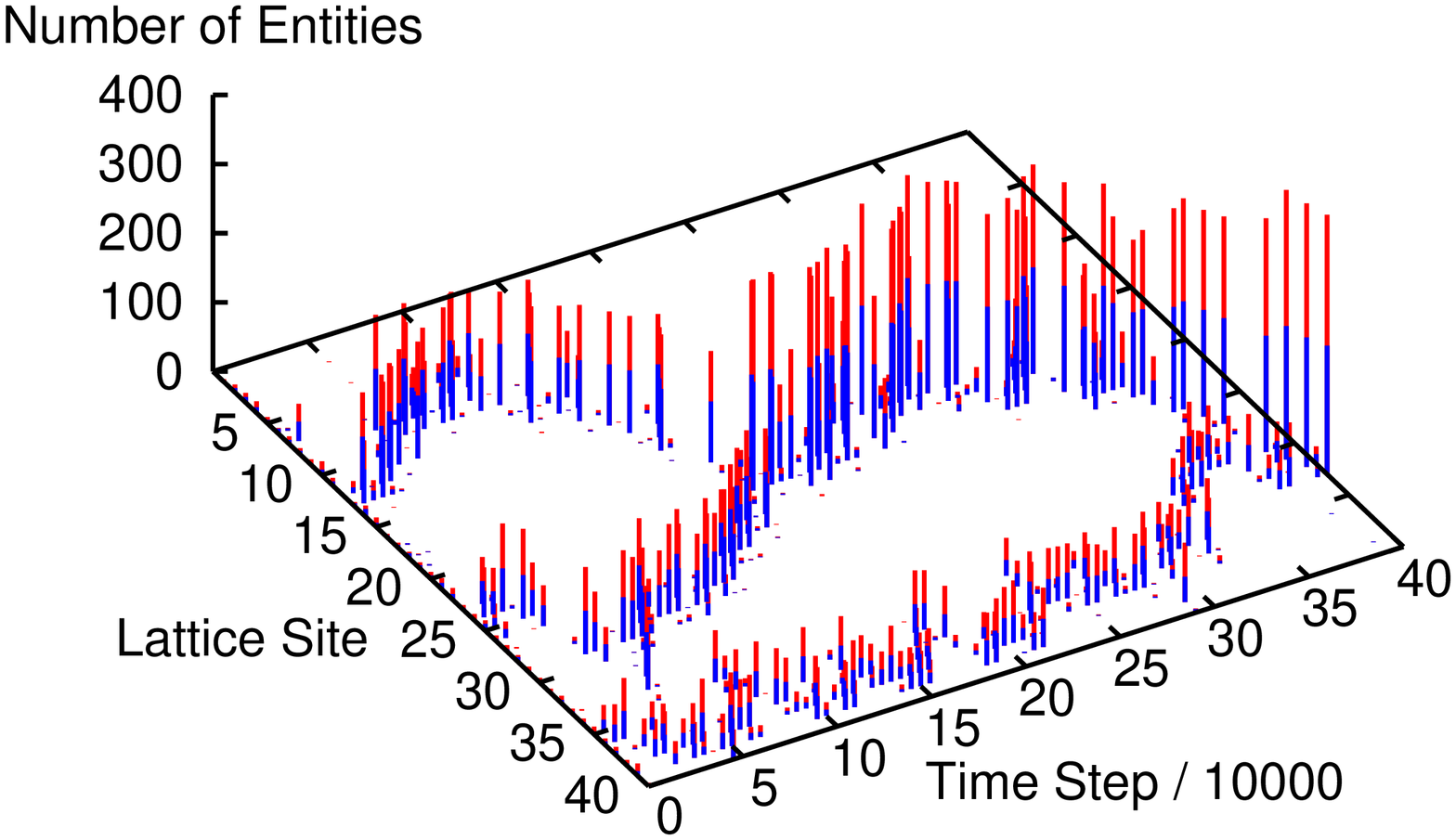}}
\end{picture} 
\end{center}
\caption[]{Illustration of the temporal evolution of the
distribution of the number of entities over the lattice sites
related to the simulation displayed in Figure~\ref{fig3}f. 
During the first few thousand time steps (not displayed), 
the small fluctuations play a subdominant role, and
the entities agglomerate around several lattice sites 
due to the assumed attractive interactions among the entities, similar
to Figure~\ref{fig3}d. 
The clusters resulting in this first dynamic stage
are more distant from each other than the range of interaction. 
Hence, their merging at later times is mainly due to fluctuations. 
The fluctuations of the individual entities around the centers of the
clusters, which originate from errors in the estimation of success,
can sum up and cause a slow variation of the centers of the clusters themselves
(see above).
In this way, initially distant clusters can accidentally come close to 
each other in the course of time, and merge. This is related with 
``evolutionary jumps''
in the overall expected success (see Figure~\ref{fig3}f) (which may
be compared to ``synergy effects'' connected with the fusion of
companies).\label{fig4}}
\end{figure}
\vfill
\clearpage
\end{document}